\documentclass[12pt,a4paper]{article}
\usepackage{graphicx}
\usepackage{amsmath}

\input{tcilatex}

\begin{document}

\title{Trapping, anomalous transport and quasi-coherent structures in magnetically
confined plasmas}
\author{{\small Madalina Vlad and Florin Spineanu} \\
{\small National Institute for Laser, Plasma and Radiation Physics, }\\
{\small Association Euratom-MEdC, P.O.Box MG-36, Magurele, Bucharest, Romania%
}}
\date{}
\maketitle

\begin{abstract}
Strong electrostatic turbulence in magnetically confined plasmas is
characterized by trapping or eddying of  particle trajectories produced by
the $E\times B$ stochastic drift. Trapping is shown to produce strong
effects on test particles and on test modes. It determines non-standard
statistics of trajectories: non-Gaussian distribution, memory effects and
coherence. Trapped trajectories form quasi-coherent structure.  Trajectory
trapping has strong nonlinear effects on the test modes on turbulent
plasmas. We determine the growth rate of drift modes as function of the
statistical characteristics of the background turbulence. We show that
trapping provides the physical mechanism for the inverse cascade observed in
drift turbulence and for the zonal flow generation.


\end{abstract}

\section{Introduction}

A component of particle motion in magnetized plasmas is the stochastic
electric drift produced by the electric field of the turbulence and by the
confining magnetic field. This drift determines a trapping effect or eddy
motion in turbulence with slow time variation \cite{kraichnan}. Typical
particle trajectories show sequences of trapping events (trajectory winding
on almost closed paths) and long jumps. Numerical simulations have shown
that the trapping process completely changes the statistical properties of
the trajectories. Particle motion in a stochastic potential was extensively
studied \cite{mccomb}-\cite{Bcarte}, but the process of trapping was not
described until recently.

New statistical methods were developed \cite{V98}, \cite{VS04} that
permitted to determine the effects of trapping. These are semi-analytical
methods based on\ a set of deterministic trajectories obtained from the
Eulerian correlation of the stochastic velocity. It was shown that trapping
determines memory effects, quasi-coherent behavior and non-Gaussian
distribution \cite{VS04}. The trapped trajectories have quasi-coherent
behavior and they form structures similar to fluid vortices. The diffusion
coefficients decrease due to trapping and their scaling in the parameters of
the stochastic field is modified. We have shown that anomalous diffusion
appears due to collisions and average flows. A review of \ the effects of
trapping on test particle statistics and on turbulent transport is presented
in the first part of this paper.

The effects of trajectory trapping on the nonlinear dynamics of the test
modes for the drift turbulence are presented in the second part of the
paper. The semi-analytical methods developed for test particles are extended
to test mode evolution in a turbulent magnetized plasmas. Test modes are
usually studied for modelling wave-wave interaction in turbulent plasmas 
\cite{K02}. A different perspective is developed here by considering test
modes on turbulent plasmas. They are described by nonlinear equations with
the advection term containing the stochastic $\mathbf{E}\times \mathbf{B}$
drift described by a stochastic field with known statistical
characteristics. The growth rate of the test modes is determined as function
of these statistical parameters. We develop a Lagrangian approach of the
type of that introduced by Dupree \cite{D66}, \cite{D72}. The difference is
that in Dupree's method the stochastic trapping of trajectory was neglected
and consequently the results can be applied to quasilinear turbulence. Our
method takes into account the trapping and the non-standard statistics of
trajectories that it yields and thus it is able to describe the nonlinear
effects appearing in strong turbulence.

The paper is organized as follows. The test particle model is presented in
Section 2. Section 3 contains and a short description of the statistical
methods. The nonlinear effects of trajectory trapping on test particle
statistics and transport are presented in Section 4. The general physical
explanation for the anomalous diffusion regimes appearing in the presence of
trajectory trapping and the formation of trajectory structures are discussed
in this section. The problem of test modes in turbulent plasmas for the case
of drift turbulence is presented in Section 5 where the growth rate and the
frequency are determined as function of the statistical characteristics of
the turbulence. The complex effects of trajectory trapping on the drift
modes are analyzed in Section 6. The conclusions are summarized in Section 7.

\section{Test particle model}

The test particle studies rely on known statistical characteristics of the
stochastic field. They are determined from experimental studies or numerical
simulations. The main aim of these studies is to determine the diffusion
coefficients. The statistics of test particle trajectories provides the
transport coefficients in turbulent plasmas without approaching the very
complicated problem of self-consistent turbulence that explains the detailed
mechanism of generation and saturation of the turbulent potential. The
possible diffusion regimes can be obtained by considering various models for
the statistics of the stochastic field.

We consider in slab geometry an electrostatic turbulence represented by an
electrostatic potential $\phi ^{e}(\mathbf{x},t),$ where $\mathbf{x}\equiv
(x_{1},x_{2})$ are the Cartesian coordinates in the plane perpendicular to
the confining magnetic field directed along $z$ axis, $\mathbf{B}=B\mathbf{e}%
_{z}$. The test particle motion in the guiding center approximation is
determined by

\begin{equation}
\frac{d\mathbf{x}(t)}{dt}=v(\mathbf{x},t)\equiv -\mathbf{\nabla }\phi (%
\mathbf{x},t)\times \mathbf{e}_{z},  \label{ec1}
\end{equation}
where $\mathbf{x}(t)$ represent the trajectory of the particle guiding
center, $\mathbf{\nabla }$ is the gradient in the $(x_{1},x_{2})$ plane and $%
\phi (\mathbf{x},t)=\phi ^{e}(\mathbf{x},t)/B$. The electrostatic potential $%
\phi (\mathbf{x},t)$ is considered to be a stationary and homogeneous
Gaussian stochastic field, with zero average. It is completely determined by
the two-point Eulerian correlation function (EC), $E(\mathbf{x},t),$ defined
by 
\begin{equation}
E(\mathbf{x},t)\equiv \left\langle \phi (\mathbf{x}^{,},t^{,})\,\phi (%
\mathbf{x}^{,}+\mathbf{x},t^{,}+t)\right\rangle .  \label{pec}
\end{equation}
The average $\left\langle ...\right\rangle $ is the statistical average over
the realizations of $\phi (\mathbf{x},t),$ or the space and time average
over $\mathbf{x}^{,}$ and $t^{,}$. This function evidences three parameters
that characterize the (isotropic) stochastic field: the amplitude $\Phi =%
\sqrt{E(\mathbf{0},0)}$, the correlation time $\tau _{c},$ which is the
decay time of the Eulerian correlation and the correlation length $\lambda
_{c},$ which is the characteristic decay distance. These three parameters
combine in a dimensionless Kubo number 
\begin{equation}
K=\tau _{c}/\tau _{fl}  \label{K}
\end{equation}
where $\tau _{fl}=\lambda _{c}/V$\ $\ $is the time of flight of the
particles over the correlation length and $V=\Phi /\lambda _{c}$ is the
amplitude of the stochastic velocity.

The diffusion coefficient is determined as (see \cite{Taylor}) 
\begin{equation}
D_{i}(t)=\int_{0}^{t}d\tau \;L_{ii}(\tau )  \label{D}
\end{equation}
where

\begin{equation}
L_{ij}(t;\,t_{1})\equiv \left\langle v_{i}(\mathbf{0},0)\,v_{j}(\mathbf{x}%
(t),t)\right\rangle  \label{CL}
\end{equation}
is the correlation of the Lagrangian velocity (LVC). It is obtained using
the decorrelation trajectory method, a semi-analytical approach presented
below.

Equation (\ref{ec1}) represents the nonlinear kernel of the test particle
problem. The statistical methods will be presented for Eq. (\ref{ec1}) for
simplicity. They were developed to include complex models with other
components of the motion (particle collisions, average flows, motion along
the confining magnetic field, etc.). The effects of these components on the
transport will be discussed in Section 4.

\section{The nested subensemble approach}

Test particle transport in magnetized plasmas in the nonlinear regime
characterized by trajectory trapping was analytically studied only the last
decade when the decorrelation trajectory method (DTM) \cite{V98} and the
nested subensemble approach (NSA) \cite{VS04} were developed. Trajectory
trapping is essentially related to the invariance of the Lagrangian
potential. Thus, a statistical method is adequate for the study of this
process if it is compatible with the invariance of the potential. The NSA is
the development of the DTM as a systematic expansion that obtains much more
statistical information.

The main idea in NSA is to study the stochastic equation (\ref{ec1}) in
subensembles of realizations of the stochastic field. First the whole set of
realizations $R$ is separated in subensembles $(S1),$ which contain all
realizations with given values of the potential and of the velocity in the
starting point of the trajectories $\mathbf{x=0}$, $t=0$: 
\begin{equation}
(S1):\quad \phi (\mathbf{0},0)=\phi ^{0},\quad \mathbf{v}(\mathbf{0},0)=%
\mathbf{v}^{0}.  \label{2}
\end{equation}
Then, each subensemble $(S1)$ is separated in subensembles $(S2)$
corresponding to fixed values of the second derivatives of the potential in $%
\mathbf{x=0}$, $t=0$ 
\begin{equation}
(S2):\quad \phi _{ij}(\mathbf{0},0)\equiv \left. \frac{\partial ^{2}\phi (%
\mathbf{x},t)}{\partial x_{i}\partial x_{j}}\right| _{\mathbf{x}=\mathbf{0}%
,t=0}=\phi _{ij}^{0}  \label{s2}
\end{equation}
where $ij=11,12,22.$ Continuing this procedure up to an order $n$, a system
of nested subensembles is constructed. The stochastic (Eulerian) potential
and velocity in a subensemble are Gaussian fields but non-stationary and
non-homogeneous, with space and time dependent averages and correlations.
The correlations are zero in $\mathbf{x=0}$, $t=0$ and increase with the
distance and time. The average potential and velocity performed in a
subensemble depend on the parameters of that subensemble and of the
subensembles that include it. They are determined by the Eulerian
correlation of the potential (see \cite{VS04} for details). The stochastic
equation (\ref{1}) is studied in each highest order subensemble $(Sn).$ The
average Eulerian velocity determines an average motion in each $(Sn).$
Neglecting the fluctuations of the trajectories, the average trajectory in $%
(Sn),$ $\mathbf{X}(t;Sn),$ is obtained from 
\begin{equation}
\frac{d\mathbf{X}(t;Sn)}{dt}=\varepsilon _{ij}\frac{\partial \Phi (\mathbf{X}%
;Sn)}{\partial X_{j}}.  \label{xmed}
\end{equation}
This approximation consists in neglecting the fluctuations of the
trajectories in the subensemble $(Sn).$ It is rather good because it is
performed in the subensemble $(Sn)$ where the trajectories are similar due
to the fact that they are super-determined. Besides the necessary and
sufficient initial condition $\mathbf{x}(0)=\mathbf{0,}$ they have
supplementary initial conditions determined by the definition (\ref{2}-\ref
{s2}) of the subensembles. The strongest condition is the inial potential $%
\phi (\mathbf{0},0)=\phi ^{0}$ that is a conserved quantity in the static
case and determines comparable sizes of the trajectories in a subensemble.
Moreover, the amplitude of the velocity fluctuations in $(Sn)$, the source
of the trajectory fluctuations, is zero in the starting point of the
trajectories and reaches the value corresponding to the whole set of
realizations only asymptotically. This reduces the differences between the
trajectories in $(Sn)$ and thus their fluctuations.

The statistics of trajectories for the whole set of realizations (in
particular the LVC) is obtained as weighted averages of these trajectories $%
\mathbf{X}(t;Sn).$ The weighting factor is the probability that a
realization belongs to the subensemble $(Sn);$ it is analytically determined.

Essentially, this method reduces the problem of determining the statistical
behavior of the stochastic trajectories to the calculation of weighted
averages of some smooth, deterministic trajectories determined from the EC
of the stochastic potential. This semi-analytical statistical approach (the
nested subensemble method) is a systematic expansion that satisfies at each
order $n>1$ all statistical conditions required by the invariance of the
Lagrangian potential in the static case. The order $n=1$ corresponds to the
decorrelation trajectory method introduced in \cite{V98}. In this case only
the average potential is conserved.

The nested subensemble method is quickly convergent. This is a consequence
of the fact that the mixing of periodic trajectories, which characterizes
this nonlinear stochastic process, is directly described at each order of
our approach. The results obtained in first order (the decorrelation
trajectory method ) for $D(t)$ are practically not modified in the second
order \cite{VS04}. Thus, the decorrelation trajectory method is a good
approximation for determining diffusion coefficients. The second order
nested subensemble method is important because it provides detailed
statistical information on trajectories: the probability of the
displacements and of the distance between neighboring trajectories in the
whole ensemble of realizations and also in the subensembles $(S1).$ A high
degree of coherence is so evidenced in the stochastic motion of trapped
trajectories.

\section{Trapping effects on test particles}

\subsection{\protect\bigskip Trajectory structures}

Detailed statistical information about particle trajectories was obtained
using the nested subensemble method \cite{VS04}. This method determines the
statistics of the trajectories that start in points with given values of the
potential. This permits to evidence the high degree of coherence of the
trapped trajectories.

The trapped trajectories correspond to large absolute values of the initial
potential while the trajectories starting from points with the potential
close to zero perform long displacements before decorrelation. These two
types of trajectories have completely different statistical characteristics 
\cite{VS04}. The trapped trajectories have a quasi-coherent behavior. Their
average displacement, dispersion and probability distribution function
saturate in a time $\tau _{s}$. The time evolution of the square distance
between two trajectories is very slow showing that neighboring particles
have a coherent motion for a long time, much longer than $\tau _{s}$. They
are characterized by a strong clump effect with the increase of the average
square distance that is slower than the Richardson law. These trajectories
form structures, which are similar with fluid vortices and represent eddying
regions. The size and the built-up time of the structures depend on the
value of the initial potential. Trajectory structures appear with all sizes,
but their characteristic formation time increases with the size. These
structures or eddying regions are permanent in static stochastic potentials.
The saturation time $\tau _{s}$ represents the average time necessary for
the formation of the structure. In time dependent potentials the structures
with $\tau _{s}>\tau _{c}$ are destroyed and the corresponding trajectories
contribute to the diffusion process. These free trajectories have a
continuously growing average displacement and dispersion. They have
incoherent behavior and the clump effect is absent. The probability
distribution functions for both types of trajectories are non-Gaussian.

The average size of the structures $S(K)$ in a time dependent potential is
plotted in Figure 1. One can see that for $K<1$ the structures are absent ($%
S\cong 0$) and that they appear for $K>1$ and continuously grow as $K$
increases. The dependence on $K$ is a power low with the exponent dependent
on the EC of the potential. The exponent is 0.19 for the Gaussian EC and
0.35 for a large EC that decays as $1/r^{2}.$

\begin{center}
\resizebox{3.7in}{!}{\includegraphics{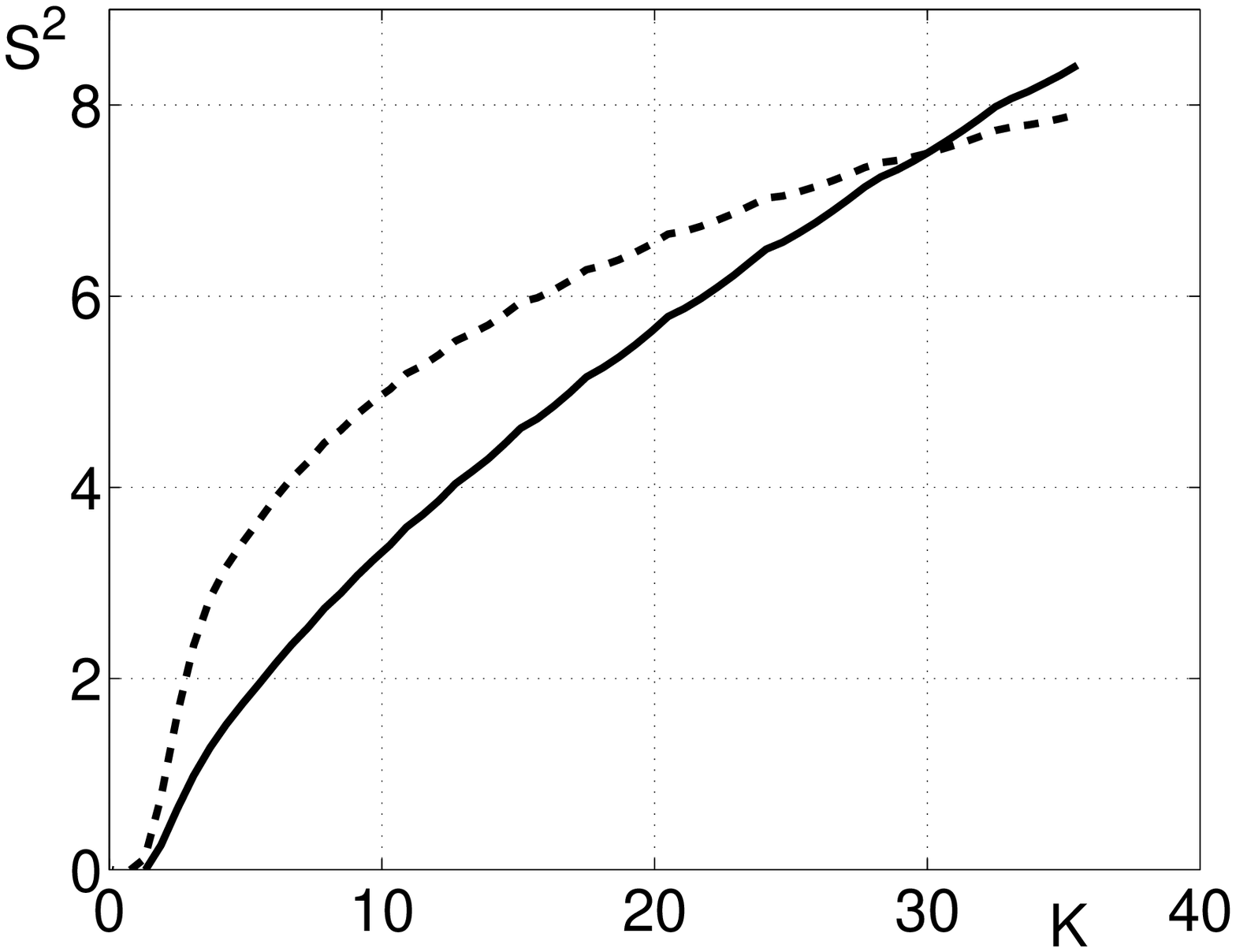}}

{\small Figure 1: The average size of the trajectory structures for Gaussian
EC (dashed line) and for an EC that decays as $1/r^{2}$ (continuous line). }
\end{center}

\subsection{Anomalous diffusion regimes}

Test particle studies connected with experimental measurements of the
statistical properties of the turbulence provide the transport coefficients
with the condition that there is space-time scale separation between the
fluctuations and the average quantities. Particle density advected by the
stochastic $\mathbf{E\times B}$ drift in turbulent plasmas leads in these
conditions to a diffusion equation for the average density with the
diffusion coefficient given by the asymptotic value of Eq. (\ref{D}). Recent
numerical simulations \cite{BJ03} confirm a close agreement between the
diffusion coefficient obtained from the density flux and the test particle
diffusion coefficient. Experiment based studies of test particle transport
permit to strongly simplify the complicated self-consistent problem of
turbulence and to model the transport coefficients by means of test particle
stochastic advection. The running diffusion coefficient $D(t)$ is defined as
the time derivative of the mean square displacement of test particles and is
determined according to Eq. (\ref{D}) as the time integral of the Lagrangian
velocity correlation (LVC). Thus, test particle approach is based on the
evaluation of the LVC for given EC of the fluctuating potential.

The turbulent transport in magnetized plasmas is a strongly nonlinear
process. It is characterized by the trapping of the trajectories, which
determines a strong influence on the transport coefficient and on the
statistical characteristics of the trajectories. The transport induced by
the $\mathbf{E\times B}$ stochastic drift in electrostatic turbulence \cite
{V04} (including effects of collisions \cite{V00}, average flows \cite{V01},
motion along magnetic field \cite{V02}, effect of magnetic shear \cite{P07})
and the transport in magnetic turbulence \cite{V03}, \cite{N04} were studied
in a series of papers using the decorrelation trajectory method. It was also
shown that a direct transport (an average velocity) appears in turbulent
magnetized plasmas due to the inhomogeneity of the magnetic field \cite{V06}%
- \cite{V08}. This statistical method was developed for the study of complex
processes as the zonal flow generation \cite{B03}, \cite{B05}.

The results of all these studies are rather unexpected when the nonlinear
effects are strong. The diffusion coefficients are completely different of
those obtained in quasilinear conditions. A rich class of anomalous
diffusion regimes is obtained for which the dependence on the parameters is
completely different compared to the scaling obtained in quasilinear
turbulence. All the components of particle motion (parallel motion,
collisions, average flows, etc.) have strong influence on the diffusion
coefficients in the non-linear regimes characterized by the presence of
trajectory trapping.

The reason for these anomalous transport regimes can be understood by
analyzing the shape of the correlation of the Lagrangian velocity for
particles moving by the $\mathbf{E}\times \mathbf{B}$ drift in a static
potential \cite{VPS}. In the absence of trapping, the typical LVC for a
static field is a function that decay to zero in a time of the order $\tau
_{fl}=\lambda _{c}/V$. This leads to Bohm type asymptotic diffusion
coefficients $D_{B}=V^{2}\tau _{fl}=V\lambda _{c}$. Only a constant $c$ is
influenced by the EC of the stochastic field and the diffusion coefficient
is $D=cD_{B}$ for all EC's. In the case of the $\mathbf{E\times B}$ drift, a
completely different shape of the LVC is obtained for static potentials due
to trajectory trapping. A typical example of the LVC is presented in Figure
2. This function decays to zero in a time of the order $\tau _{fl}$ but at
later times it becomes negative, it reaches a minimum and then it decays to
zero having a long, negative tail. The tail has power law decay with an
exponent that depends on the EC of the potential \cite{V04}. The positive
and negative parts compensate such that the integral of $L(t)$, the running
diffusion coefficient $D(t)$, decays to zero. The transport in static
potential is thus subdiffusive. The long time tail of the LVC shows that the
stochastic trajectories in static potential have a long time memory.

This stochastic process is unstable in the sense that any weak perturbation
produces a strong influence on the transport. A perturbation represents a
decorrelation mechanism and its strength is characterized by a decorrelation
time $\tau _{d}$. The weak perturbations correspond to long decorrelation
times, $\tau _{d}>\tau _{fl}.$ In the absence of trapping, such a weak
perturbation does not produce a modification of the diffusion coefficient
because the LVC is zero at $t>\tau _{fl}.$ In the presence of trapping,
which is characterized by long time LVC as in Figure 2, such perturbation
influences the tail of the LVC and destroys the equilibrium between the
positive and the negative parts. Consequently, the diffusion coefficient is
a \textit{decreasing function of }$\tau _{d}.$ It means that when the
decorrelation mechanism becomes stronger ($\tau _{d}$ decreases) the
transport increases. This is a consequence of the fact that the long time
LVC is negative. This behavior is completely different of that obtained in
stochastic fields that do not produce trapping. In this case, the transport
is stable to the weak perturbations. An influence of the decorrelation can
appear only when the later is strong such that $\tau _{d}<\tau _{fl}$ and it
determines the increase of the diffusion coefficient with the increase of $%
\tau _{d}$. This inverse behavior appearing in the presence of trapping is
determined by the fact that a stronger perturbation (with smaller $\tau _{d}$%
) liberates a larger number of trajectories, which contribute to the
diffusion.

\begin{center}
\resizebox{3.7in}{!}{\includegraphics{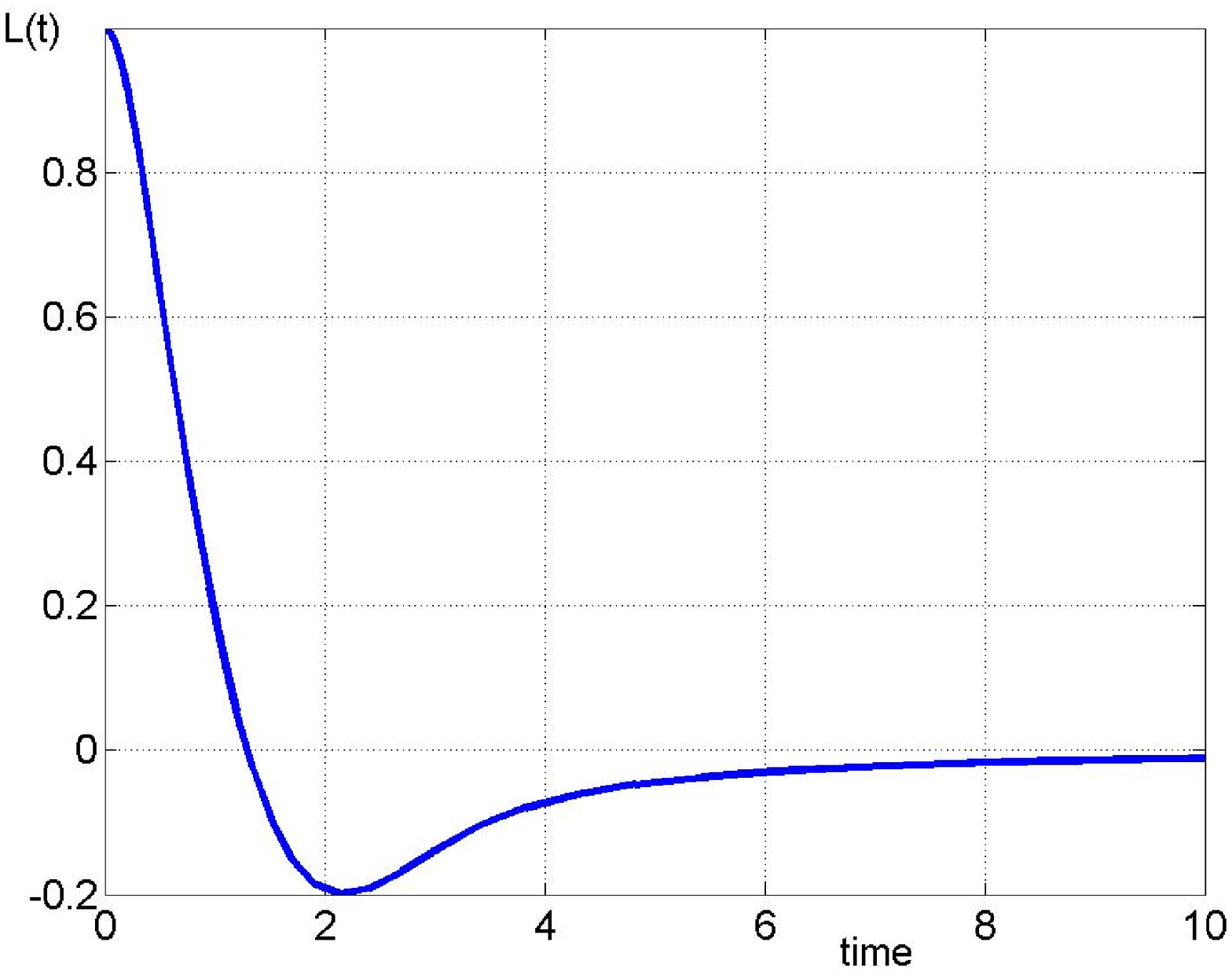}}

{\small Figure 2: Typical Lagrangian velocity correlation in static
potential.}
\end{center}

\vspace*{0.1in}

The decorrelation can be produced for instance by the time variation of the
stochastic potential, which produces the decay of both Eulerian and
Lagrangian correlations after the correlation time $\tau _{c}$. The
decorrelation time in this case is $\tau _{c}$ and it is usually represented
by a dimensionless parameter, the Kubo number defined by Eq. (\ref{K}). The
transport becomes diffusive with an asymptotic diffusion coefficient that
scales as $D_{tr}=cV\lambda _{c}K^{\gamma }$, with $\gamma $ in the interval
[-1, 0] (trapping scaling \cite{V04}). The diffusion coefficient is a
decreasing function of $\tau _{c}$ in the nonlinear regime $K>1$.

For other types of perturbations, their interaction with the trapping
process produces more complicated nonlinear effects. For instance, particle
collisions lead to the generation of a positive bump on the tail of the LVC 
\cite{V00} due to the property of the 2-dimensional Brownian motion of
returning in the already visited places. Other decorrelation mechanisms
appearing in plasmas are average component of the velocity like poloidal
rotation \cite{V01} or the parallel motion that determines decorrelation
when the potential has a finite correlation length along the confining
magnetic field. The effects of an average component of the velocity are
discussed in Section 5.1. in connection with drift turbulence.

\section{Test modes on drift turbulence}

Test particle trajectories are strongly related to plasma turbulence. The
dynamics of the plasma basically results from the Vlasov-Maxwell system of
equations, which represents the conservation laws for the distribution
functions along particle trajectories. Studies of plasma turbulence based on
trajectories were initiated by Dupree \cite{D66}, \cite{D72} and developed
especially in the years seventies (see the review paper \cite{K02} and
references there in). These methods do not account for trajectory trapping
and thus they apply to the quasilinear regime or to unmagnetized plasmas. A
very important problem that has to be understood is the effect of the
non-standard statistical characteristics of the test particle trajectories
on the evolution of the instabilities and of turbulence in magnetized
plasmas.

We extend the Lagrangian methods of the type of \cite{D72}, \cite{SV88}, 
\cite{VSM} to the nonlinear regime characterized by trapping. We study
linear modes on turbulent plasma with the statistical characteristics of the
turbulence considered known. The dispersion relation for such test modes is
determined as function of the characteristics of the turbulence. We consider
the drift instability in slab geometry with constant magnetic field. The
combined effect of the parallel motion of electrons (non-adiabatic response)
and finite Larmor radius of the ions destabilizes the drift waves.

The gyrokinetic equations are not linearized around the unperturbed state as
in the linear theory but around a turbulent state with known spectrum. The
perturbations of the electron and ion distribution functions are obtained
from the gyrokinetic equation using the characteristics method as integrals
along test particle trajectories of the source terms determined by the
average density gradient.

The background turbulence produces two modifications of the equation for the
linear modes. One consists in the stochastic $\mathbf{E\times B}$ drift that
appears in the trajectories and the other is the fluctuation of the
diamagnetic velocity. Both effects are important for ions while the response
of the electrons is approximately the same as in quiescent plasma. They
depend on the parameters of the turbulence.

\subsection{The statistics of the characteristics}

The solution for the potential in the zero Larmor radius limit is 
\begin{equation}
\phi (\mathbf{x},z,t)=\phi _{0}(\mathbf{x-V}_{\ast }t,z),  \label{s0}
\end{equation}
where $\phi _{0}$ is the initial condition and $\mathbf{V}_{\ast }$ is the
diamagnetic velocity. This shows that the potential is not changed but
displaced with the diamagnetic velocity. The finite Larmor radius effects
consist in the modification of the amplitude and of the shape of the
potential, but this appears on a much slower time scale.

The ordering of the characteristic times for the drift turbulence is 
\begin{equation}
\tau _{\parallel }^{e}\ll \tau _{\ast }\ll \tau _{c}\ll \tau _{\parallel
}^{i},  \label{ord}
\end{equation}
where $\tau _{\parallel }^{e},\tau _{\parallel }^{i}$ are the parallel
decorrelation times for electrons and ions ($\tau _{\parallel
}^{e,i}=\lambda _{\parallel }/v_{th}^{e,i}$ with $\lambda _{\parallel }$\
the parallel correlation length and $v_{th}^{e,i}$ the thermal velocity), $%
\tau _{\ast }=\lambda _{c}/V_{\ast }$ is the characteristic time for the
potential drift and $\tau _{c}$\ is the correlation time of the potential.
The linear and nonlinear regimes are determined by the position of the time
of flight in this ordering. The latter is much smaller than $\tau _{c}$ and
much larger than $\tau _{\parallel }^{e}.$ The statistical characteristics
of the trajectories essentially depend on the ratio $\tau _{\ast }/\tau
_{fl}.$

The quasilinear case corresponds to $\tau _{\ast }/\tau _{fl}\ll 1$ ($%
V/V_{\ast }\ll 1),$ which means turbulence with the amplitude of the $%
\mathbf{E}\times \mathbf{B}$ drift smaller than the diamagnetic velocity.
The motion of the potential produces in this case a fast decorrelation and
trapping does not appear. The probability of displacements is Gaussian and
the diffusion coefficient is $D_{ql}=V^{2}\tau _{\ast }.$

The nonlinear case corresponds to $\tau _{\ast }/\tau _{fl}>1$ ($V/V_{\ast
}>1).$ The motion of the potential is slow and trajectory structures
produced by trapping exist in this case.

The test particle motion in a drifting potential is obtained by a Galilean
transformation from the motion produced by a stochastic $\mathbf{E}\times 
\mathbf{B}$ drift and an average velocity $V_{d}.$ This process was studied
in \cite{V03}. It was shown that strips of opened contour lines of the
effective potential $\phi +xV_{d}$ appear due to an average velocity $V_{d}$
and that the width of these strips increases with $V_{d}$ until they
completely eliminate the closed contour lines (for $V_{d}>V$). The
Lagrangian correlation of the velocity in the presence of an average
velocity $V_{d}<V$ does not decay to zero as in Figure 2, but it has a
positive asymptotic values at $t\rightarrow \infty .$ Consequently the
transport along the average velocity is superdiffusive in the static
potential and diffusive with large diffusion coefficient (proportional with
the average velocity) in the time dependent case. A part of the particles
are trapped and the other move on the strips of opened contour lines of the
effective potential. The invariance of the distribution of the Lagrangian
velocity shows that the average velocity of the free particles $%
V_{fr}^{\prime }$ fulfils the condition 
\begin{equation}
n_{fr}V_{fr}^{\prime }=V_{d},  \label{vcons}
\end{equation}
and thus it is larger than the average velocity ($V_{fr}^{\prime }>V_{d}).$ $%
n_{tr}$ is the fraction of trapped trajectories and $n_{fr}$\ is the
fraction of free trajectories at a moment ($n_{tr}+n_{fr}=1)$.

This physical image leads, by changing the reference frame, to the following
paradigm of the statistics of trajectories produced by the $\mathbf{E}\times 
\mathbf{B}$ drift in a moving potential. The trapped particles (structures)
are advected by the moving potential while the other particles have an
average motion in the opposite direction with a velocity $V_{fr}$ such that 
\begin{equation}
n_{fr}V_{fr}+n_{tr}V_{\ast }=0,  \label{vcons2}
\end{equation}
which is the equivalent of Eq. (\ref{vcons}). This shows that there are
particle flows in opposite directions induced by the drifting potential if
the amplitude of the stochastic $\mathbf{E}\times \mathbf{B}$ velocity is
larger than the velocity of the potential. This determines a spliting of the
probability of displacements in two parts: the probability of trapped and
the probability of free particles. The first is a picked function that has
constant width and moves with the velocity $V_{\ast }.$ The second, is a
Gaussian like function with an average displacement $\left\langle
x_{2}\right\rangle _{fr}=V_{fr}t=-V_{\ast }t\;n_{tr}/n_{fr}.$ The
probability of displacements at $t<\tau _{c}$ is modeled by

\begin{equation}
P(x,y,t)=n_{tr}G(x,y-V_{\ast
}t;S_{x},S_{y})+n_{fr}G(x,y-V_{fr}t;S_{x}+2D_{x}t,S_{y}+2D_{y}t)
\label{prob}
\end{equation}
where $G(x,y;S_{x},S_{y})$ is the 2-dimensional Gaussian distribution with
dispersion $S_{x},S_{y}.$ We have considered for simplicity the distribution
of trapped particles as a Gaussian function but with small (fixed)
dispersion. The shape of this function does not change much these
estimations. The free trajectories have dispersion that grows linearly in
time.

\subsection{The growth rate of drift modes in turbulent plasma}

The average propagator of for a mode with frequency $\omega $ and wave
number $\mathbf{k}=\left( k_{1},k_{2}\right) $ is evaluated using the above
results on trajectory statistics. It depends on the size $S(K)$ of the
structures and on the fractions of trapped and free particles: 
\begin{eqnarray}
&&\int_{-\infty }^{t}d\tau \left\langle \exp \left( -i\mathbf{k}\cdot 
\mathbf{x}(\tau )\right) \right\rangle \exp \left( i\omega (t-\tau )\right) 
\notag \\
&=&i\exp \left( -k_{i}^{2}S_{i}^{2}\right) \left[ \frac{n_{tr}}{\omega
+k_{y}V_{\ast }}+\frac{n_{fr}}{\omega +k_{y}V_{fr}+ik_{i}^{2}D_{i}}\right]
\label{prop}
\end{eqnarray}
where $\mathbf{x}(\tau )$ is the trajectory in the moving potential
integrated backward in time with the condition $\mathbf{x}$ at time $t.$

The solution of the dispersion relation is obtained as 
\begin{equation}
\omega =k_{y}V_{\ast }^{eff}  \label{omeg}
\end{equation}
\begin{equation}
V_{\ast }^{eff}=V_{\ast }\frac{\Gamma _{0}\mathcal{F(}n_{fr}-n_{tr})+2n_{tr}%
}{2-\Gamma _{0}\mathcal{F}}  \label{veff}
\end{equation}
\begin{equation}
\mathcal{F}=\exp \left( -\frac{1}{2}k_{i}^{2}S^{2}\right)  \label{fstr}
\end{equation}
\begin{equation}
\gamma =\frac{\sqrt{\pi }}{\left| k_{z}\right| v_{Te}}\frac{k_{y}^{2}\left(
V_{\ast }-V_{\ast }^{eff}\right) \left( V_{\ast }^{eff}-\frac{n_{tr}}{n_{fr}}%
V_{\ast }\right) }{2-\Gamma _{0}\mathcal{F}}-k_{i}^{2}D_{i}\frac{2-\Gamma
_{0}\mathcal{F}n_{tr}}{2-\Gamma _{0}\mathcal{F}}+k_{i}k_{j}R_{ij}V_{\ast
}^{eff}  \label{gam}
\end{equation}
where $\Gamma _{0}=exp(-b)I_{0}(b),$ $b=k_{\perp }^{2}\rho _{L}^{2}/2$\ \
and $\rho _{L}$ is the ion Larmor radius. The tensor $R_{ij}$ has the
dimension of a length and is defined by 
\begin{equation}
R_{ji}(\tau ,t)\equiv \int_{\tau }^{t}d\theta ^{\prime }\int_{-\infty
}^{\tau -\theta ^{\prime }}d\theta M_{ji}(\left| \theta \right| )
\label{rji}
\end{equation}
where $M_{ij}$ is the Lagrangian correlation 
\begin{equation}
M_{ji}(\left| \theta ^{\prime }-\theta \right| )\equiv \left\langle
v_{j}\left( \mathbf{x}^{i}(\theta ^{\prime }),z,\theta ^{\prime }\right)
\;\partial _{2}v_{i}\left( \mathbf{x}^{i}(\theta ),z,\theta \right)
\right\rangle ,  \label{mji}
\end{equation}
and $v_{j}$ is the $\mathbf{E\times B}$ drift velocity.

Several effects appear in the test modes characteristics due to the
background turbulence. The spreading of ion trajectories produces the
diffusion $D_{i}$ that influences the growth rate (\ref{gam}) both in linear
and nonlinear conditions. This term is similar to the result of Dupree, but
the values of \ $D_{i}$ is influenced by trapping. Beside this, there are
several influences that appear only in the nonlinear regime. The first is
the factor $\mathcal{F}$ given by Eq. (\ref{fstr}), which is produced by the
trajectory structures. It determines essentially the modification of the
mode frequency. The flows of the ions induced by the drifting potential are
represented by the fractions $n_{tr}$ and $n_{fr}.$The tensor $R_{ij}$ is
determined by the fluctuations of the diamagnetic velocity due to the
background turbulence. We will analyze each of these processes in the next
section.

\section{\protect\bigskip Trapping effects on the test modes}

The trajectory trapping has a complex influence on the mode. This can be
understood by considering the evolution of the drift turbulence starting
from a stochastic potential with very small amplitude as it can be deduced
from the growth rates of the test modes.

The trajectories are Gaussian, there is no trapping in such potential and
the only effect of the background turbulence is the diffusion of ion
trajectories that produce resonance broadening. The well known results of
drift modes in quasilinear turbulence are obtained 
\begin{equation}
\omega =k_{y}V_{\ast }\frac{\Gamma _{0}}{2-\Gamma _{0}},\quad \gamma =\frac{%
\sqrt{\pi }}{\left| k_{z}\right| v_{Te}}\frac{\left( k_{y}V_{\ast }-\omega
\right) \left( \omega -k_{y}V_{\ast }\right) }{2-\Gamma _{0}}%
-k_{i}^{2}D_{ql},  \label{ql}
\end{equation}
where $D_{x}=D_{y}=D_{ql}=V^{2}\lambda _{c}/V_{\ast }.$\ This shows that the
modes with large $k$ are damped due to ion trajectory diffusion as the
amplitude potential increases. The maximum of the spectrum is for $\omega
=k_{y}V_{\ast }/2$ and corresponds to $k_{\perp }\rho _{L}\sim 1.$

When the nonlinear stage is attaint for $V>V_{\ast },$ the first effect is
produced when the fraction of trapped trajectories is still small by the
quasi-coherent component of ion motion. The structures of ion trajectories
determine the $\mathcal{F}$ factor (\ref{fstr}), which modifies the
effective diamagnetic frequency (\ref{veff}) and the frequency $\omega $. At
this stage the flows can be neglected ($n_{tr}\cong 0,$ $n_{fr}\cong 1)$ and 
$R_{ji}\cong 0$ in Eqs. (\ref{omeg})-(\ref{gam}), so that only the factor $%
\mathcal{F}$ is important. It is interesting to note that this factors
appears in Eqs. (\ref{omeg})-(\ref{gam}) as a multiple of $\Gamma _{0},$
although they come from different sources ($\mathcal{F}$ from the propagator
and $\Gamma _{0}$ from gyro-average of the mode potential). This shows that
the trapping or eddying motion has the same attenuation effect as the
gyro-average. The maximum of the spectrum that appears for $\omega
=k_{y}V_{\ast }/2$ is obtained for smaller $k_{\perp },$ of the order of the
size of the trajectory structures $k_{\perp }S\sim 1.$ This means that the
unstable range of the wave numbers is displaced toward small values. The
maximum growth rate is not changed but displaced at values of the order $%
1/S. $ Consequently, in this stage both the amplitude of the turbulence and
its correlation length increase.

At larger amplitude of the background potential, when the fraction or
trapped ions becomes comparable with the fraction of free ions, the ion
flows induced by the moving potential become important. These flows
determine the increase of the effective diamagnetic velocity (\ref{veff})
toward the diamagnetic velocity and the modification of the growth rate of
the drift modes. The latter decreases and for $n_{tr}=n_{fr}$ it is
negative. The evolution of the amplitude becomes slower and eventually the
growth rates vanishes and changes the sign. Thus, the flows of the ions
induced by the moving potential produce the damping of the drift modes.

The fluctuations of the diamagnetic velocity due to background turbulence
determine a direct contribution to the growth rate (the tensor $R_{ij}).$
This term is zero for homogeneous and izotropic turbulence and strongly
depends on the parameters of the anizotropy. The $i=j=1$ component
corresponds to zonal flows. Preliminary results show that it appears for
trapped particles due to the anizotropy induced by ion flows with the moving
potential.

\section{Summary and conclusions}

We have discussed the problem of stochastic advection of test particles by
the $\mathbf{E\times B}$ drift in turbulent plasmas. We have shown that
trajectory trapping or eddying have complex nonlinear effects on the
statistical characteristics of the trajectories and on the transport. The
nonlinear effects are very strong in the case of static potentials. The
trajectories are non-Gaussian, there is statistical memory, coherence and
they form structures. These properties persist if the system is weakly
perturbed by time variation of the potential or by other components of the
motion (collisions, poloidal rotation, parallel motion). The memory effect
(long tail of the LVC) determines anomalous diffusion regimes.

The process of trajectory trapping also influences the evolution of the
turbulence. Recent results on test modes on turbulent plasmas are presented.
They are based on a Lagrangian method that takes into account the trapping
or eddying of the ions. The growth rate and the frequency of the drift modes
on turbulent plasmas are estimated as function of the characteristics of the
turbulence. The effects of the background turbulence appear in particle
trajectories (characteristics of Vlasov equations) and in the fluctuations
of the diamagnetic velocity produced by the density fluctuations. We show
that the nonlinear process of trapping, which determines non-standard
statistical properties of trajectories, has a very strong and complex
influence on the evolution of the turbulence. It appears when the amplitude
of the $\mathbf{E}\times \mathbf{B}$ drift becomes larger than the
diamagnetic velocity.

A different physical perspective on the nonlinear evolution of drift waves
is obtained. The main role is played by the trapping of the ions in the
stochastic potential that moves with the diamagnetic velocity. We show that
the moving potential determines flows of the ions when the amplitude of the $%
\mathbf{E}\times \mathbf{B}$ velocity is larger than the diamagnetic
velocity. A part of the ions are trapped and move with the potential while
the other ions drift in the opposite direction. These opposite (zonal) flows
compensate such that the average velocity is zero. The evolution of the
turbulence toward large wave lengths (the inverse cascade) is determined by
ion trapping, which averages the potential and determines a smaller
effective diamagnetic velocity. The ion flows produced by the moving
potential determine the decay of the growth rate and eventually the damping
of the drift modes and generate zonal flows due to their nonlinear
interaction with the fluctuations of the diamagnetic velocity.

\end{document}